\documentclass[12pt]{article}
\usepackage{epsfig,amssymb}
\usepackage{latexsym}

\newcommand{\bq}{\begin{equation}}
\newcommand{\eq}{\end{equation}}
\newcommand{\bqa}{\begin{eqnarray}}
\newcommand{\eqa}{\end{eqnarray}}
\newcommand{\ben}{\begin{enumerate}}
\newcommand{\een}{\end{enumerate}}
\newcommand{\bc}{\begin{center}}
\newcommand{\ec}{\end{center}}
\newcommand{\bqb}{\begin{eqnarray*}}
\newcommand{\eqb}{\end{eqnarray*}}

\def\lsim{\lesssim}

\def\pr#1#2#3{ Phys. Rev. ${\bf{#1}}$:#2 (#3)}
\def\prl#1#2#3{ Phys. Rev. Lett. ${\bf{#1}}$:#2 (#3)}
\def\pl#1#2#3{ Phys. Lett. ${\bf{#1}}$:#2 (#3)}
\def\prep#1#2#3{ Phys. Rept. ${\bf{#1}}$:#2 (#3)}
\def\rmp#1#2#3{ Rev. Mod. Phys. ${\bf{#1}}$:#2(#3)}

\def\np#1#2#3{ Nucl. Phys. ${\bf{#1}}$:#2 (#3)}
\def\zp#1#2#3{ Z. f. Phys. ${\bf{#1}}$:#2 (#3)}

\def\epj#1#2#3{ Eur. Phys. J. ${\bf{#1}}$:#2 (#3)}

\def\app#1#2#3{Astropart. Phys. ${\bf{#1}}$:#2 (#3)}
\def\aop#1#2#3{Annals of Phys. ${\bf{#1}}$:#2 (#3)}

\def\polon#1#2#3{Acta Phys. Polon. ${\bf{#1}}$:#2 (#3) }

\def\eg{{\it e.g. }}

\def\etal{{\it et.al.}}

\global\nulldelimiterspace = 0pt

\def\Ocal{{\cal O}}

\def\swd{s^2_W}

\def\mw{m_W}
\def\mz{m_Z}
\def\mzd{m_Z^2}
\def\mi{m_i}
\def\mj{m_j}
\def\mid{m_i^2}
\def\mjd{m_j^2}

\def\tchi{\tilde \chi^0}

\begin{document}
\pagenumbering{arabic}
\thispagestyle{empty}
\def\thefootnote{\fnsymbol{footnote}}
\setcounter{footnote}{1}

\begin{flushright}
March  2006, corrected version\\
PTA/06-04\\
hep-ph/0602049.\\

 \end{flushright}
\vspace{2cm}
\begin{center}
{\Large\bf Neutralino-neutralino  annihilation    to
$\gamma Z$ in MSSM\footnote{Work supported
by the Greek Ministry of Education and Religion and the
EPEAEK program Pythagoras, and by the European Union
program   HPRN-CT-2000-00149.}}\\
 \vspace{0.5cm}
{ Th. Diakonidis$^a$,
G.J. Gounaris$^a$, J. Layssac$^b$, P.I. Porfyriadis$^a$ and  F.M. Renard$^b$}\\
\vspace{0.2cm}
$^a$Department of Theoretical Physics, Aristotle
University of Thessaloniki,\\
Gr-54124, Thessaloniki, Greece.\\
\vspace{0.2cm}
$^b$Laboratoire de Physique
Th\'{e}orique et Astroparticules, UMR 5207,\\
Universit\'{e} Montpellier II,
 F-34095 Montpellier Cedex 5.\\

\vspace*{1.cm}

{\bf Abstract}
\end{center}
 The  1-loop computation of the processes
$\tchi_i \tchi_j \to \gamma Z$ has been performed
at   an arbitrary  c.m. energy for any
pair of  MSSM neutralinos. As an  application suitable for Dark
Matter (DM) searches, the neutralino-neutralino annihilation is studied
at the   limiting case  of vanishing relative velocity, describing
the present DM distribution in the galactic halo; and
at a relative velocity of about 0.5,
  determining the neutralino relic density contributions.
 The most useful situation is obviously for $i=j=1$, but
 the case of non-identical  neutralinos
 may also be useful in some corners  of the parameter space.
Our results are contained in the FORTRAN code PLATONdmgZ, applying
 to  any set of real MSSM parameters.
Numerical results   are also presented  for a sample of 6 MSSM
models,   describing the
various possible neutralino properties. A comparison with  other
existing works is  also made.

\vspace{0.5cm}

\def\thefootnote{\arabic{footnote}}
\setcounter{footnote}{0}
\clearpage

\section{Introduction}

The nature of  Cold Dark Matter (DM), constituting almost
23\% of the energy of the Universe,
is one of the most exciting  subjects of
physics today \cite{Kamio-rep, Laz}.
Within the minimal R-parity conserving Supersymmetric (SUSY)
 framework, a  most obvious candidate  for such matter
  is of course   the lightest neutralino $\tchi_1$
 \cite{Kamio-rep}. A  striking signature for such DM
 would then be the detection of $\gamma$-rays obtained
 through the annihilation of two neutralinos \cite{SUSYsearches}.

 The spectrum of most of
 these $\gamma$-rays should be  continuous \cite{SUSYsearches}.
 But once the necessary sensitivity is reached,
  sharp $\gamma$-rays could also be observed, induced by
  neutralino-neutralino annihilation at rest, to
 $\gamma \gamma$, $\gamma Z$ or a photon together with a neutral Higgs particle.
Observing such $\gamma$-rays, with
 the predicted ratio of intensities, would be a really great discovery.

 For identical annihilating  neutralinos in the vanishing
 relative velocity limit, the processes  $\tchi_1\tchi_1 \to \gamma \gamma$ and
 $\tchi_1\tchi_1\to \gamma Z$   have already been studied  in
 \cite{Bergstrom-gg} and \cite{Bergstrom-gZ} respectively.  \par

 Subsequently,  $\tchi_i\tchi_j \to \gamma \gamma$ has also
 been studied in \cite{DMgg}, at any relative velocity
 and for any neutralino pair. This study is based on analytically
 expressing the amplitudes, in terms
 of Passarino-Veltman (PV) functions \cite{Veltman}.
 Such   non-vanishing relative velocity results
  may  be useful in estimating the
 neutralino contribution to the DM density \cite{SUSYsearches}.
 In   some, admittedly extreme  corners
of  the parameter space, where the two lightest neutralinos
may be exactly degenerate, the $(i\neq j)$-case may also be useful.

  The same analytic amplitudes were then also used
to study the reverse process $\gamma \gamma \to \tchi_i\tchi_j$
in a $\gamma\gamma$ Linear Collider ($LC_{\gamma \gamma}$) \cite{LCggnn};
while the related   amplitudes   for $gg  \to \tchi_i\tchi_j$
and $gg  \to \tchi_i \tilde g $,
were used to determine    the corresponding production of
a pair of neutralinos  \cite{LHCnn},
 or a single neutralino  \cite{LHCn}, at the CERN LHC.

 FORTRAN codes supplying all 1-loop contributions to these processes,
 for  any set of real MSSM parameters at the electroweak scale,
and any neutralino pair,  may be obtained from \cite{plato}.
 These  should be useful for  checking the consistency of the  neutralino DM
 identification, using  collider experiments. The importance
 of such consistency checks can hardly be overemphasized.

The annihilation process
 $\tchi_1\tchi_1 \to \gamma \gamma$, for the lightest neutralino,
 at any relative velocity, has also been studied  recently
  by  \cite{Boudjema}. In this reference,
 an automatic  numerical method is presented, which directly
 calculates the needed 1-loop amplitudes, starting from the Feynman diagrams.
As pointed out by \cite{Boudjema}, their results for  $\gamma \gamma$-annihilation,
perfectly agree with  those of  \cite{Bergstrom-gg} and \cite{DMgg}.

However, \cite{Boudjema} also gives results for
$\tchi_1\tchi_1 \to \gamma Z$ at  $v_{11}=0.5$ and $v_{11}=0$.
Discrepancies appear though, when comparing these
results with those of \cite{Bergstrom-gZ}, which only exist for $v_{11}= 0$.
An independent calculation of this process  seems therefore
required. As for the $\gamma \gamma$-case,
 it would  be advantageous to have  $\gamma Z$-results
for any pair  of neutralinos, at any relative velocity,
which may allow also the study of the reverse process in a
future  $e\gamma$ linear collider.

Therefore, in this paper we present an analytic calculation of the
1-loop process $\tchi_i\tchi_j \to \gamma Z$, following   the same
philosophy as in \cite{DMgg, LCggnn}. The results are based on
analytically expressing  the helicity amplitudes in terms of PV
functions, and they are  valid for  any set of real MSSM
parameters\footnote{As   usual, the vacuum  velocity of light
$c$, is taken as the unit of velocities.}.
They are contained in the  FORTRAN code PLATONdmZ,
which is herewith released in \cite{plato}.

As we will discuss below, the  results announced here
  agree with those of \cite{Boudjema}  for  i=j=1,
at vanishing relative velocities, while they   deviate from those of
\cite{Bergstrom-gZ}. In the $v_{11}=0.5$ case though, a perfect agreement  with
\cite{Boudjema} exists only for five of the six considered models, while for
the other one the agreement is only approximate.
Our procedure is presented in  Section 2 below,  the results in Section 3,
and an Outlook is given in  Section 4.\\

\section{Procedure}

The process studied  is
\bq
 \tchi_i(l_1, \lambda_1)~+~\tchi_j(l_2, \lambda_2)
\to \gamma(p_1, \mu_1)+ Z(p_2, \mu_2) ~~ , \label{process-gZ} \eq
where the momenta and  helicities of the incoming neutralinos and
the outgoing vector bosons are indicated explicitly. Generally, the
incoming neutralinos may be different. The corresponding helicity
amplitudes,  satisfying  the  standard Jacob-Wick conventions
\cite{JW}, are denoted as
$F^{ij}_{\lambda_1,\lambda_2;\mu_1,\mu_2}(\theta)$, where $\theta$
is the c.m. scattering angle of the outgoing photon with respect
to the incoming neutralino $\tchi_i$.

According to \cite{JW},  the \underline{$\chi^0_i\chi^0_j$ antisymmetry}
due to the fermionic nature of the neutralinos, obliges the helicity amplitudes
to obey
\bq
F^{ij}_{\lambda_1,\lambda_2;\mu_1,\mu_2}(\theta)=(-1)^{\mu_1-\mu_2}~
F^{ji}_{\lambda_2,\lambda_1;\mu_1,\mu_2}(\pi-\theta)~~, \label{ij-antisym}
\eq
while the \underline{CP symmetry} for real SUSY-braking parameters implies
\bq
F^{ij}_{-\lambda_1,-\lambda_2;-\mu_1,-\mu_2}(\theta)=
(-1)^{\lambda_1-\lambda_2-(\mu_1-\mu_2)}~\eta_i\eta_j
F^{ij}_{\lambda_1,\lambda_2;\mu_1,\mu_2}(\theta)~, \label{CP-sym}
\eq
where  $\eta_i=\pm 1$ is the CP eigenvalue  of $\tchi_i$ \cite{LeMouel1}.
Relation (\ref{ij-antisym}) is very  important,
since it is used below to  reduce the number of needed independent
diagrams.

In the calculation we use the 'tHooft-Feynman gauge
$\xi_W=\xi_Z=\xi_\gamma=1$, together with the   standard linear
gauge fixing.

\begin{figure}[p]
\[
\hspace{0.cm}\epsfig{file=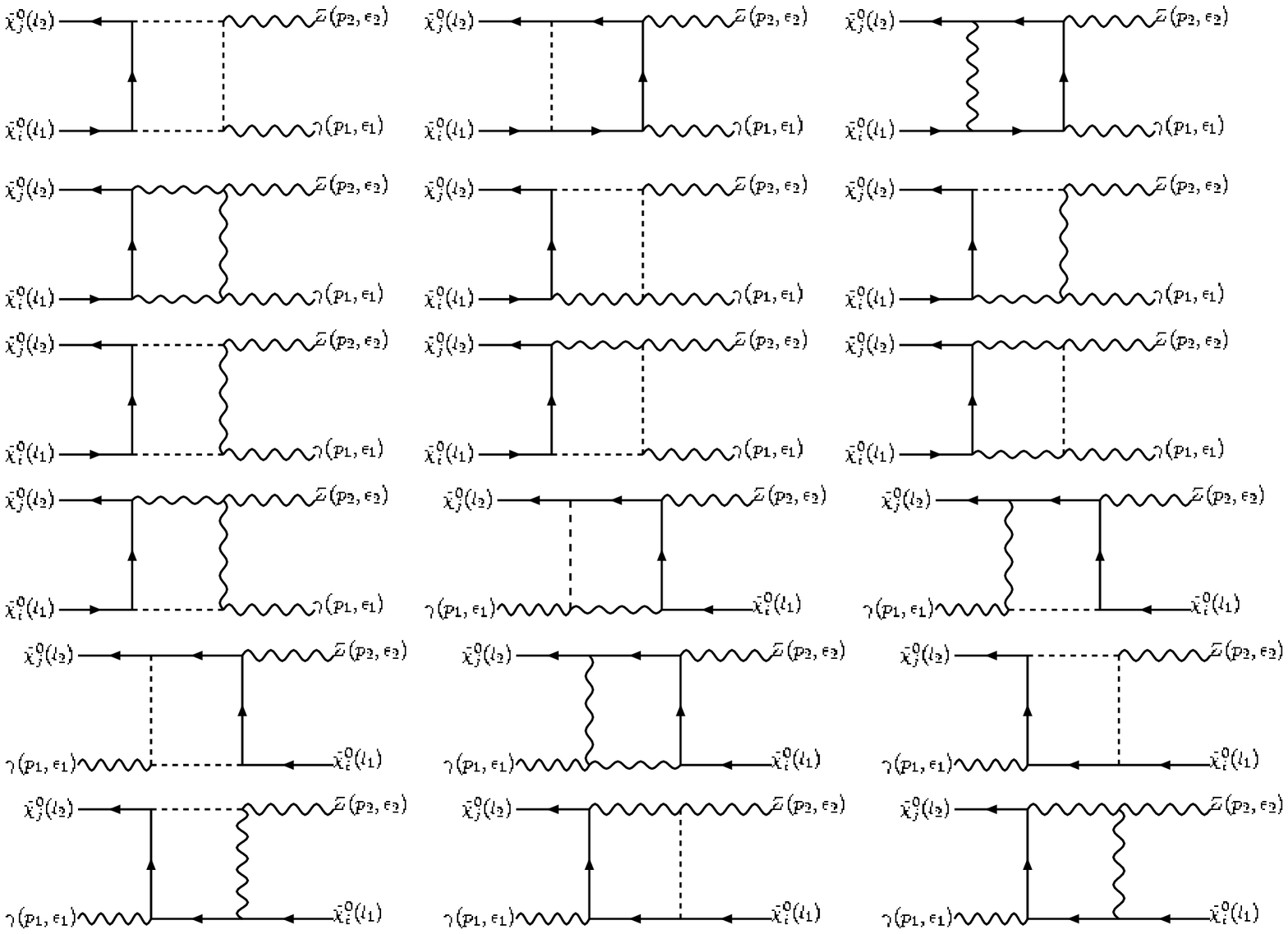,height=20.cm, width=15.cm}
\]
\vspace*{-9cm}
\caption[1]{ Box Feynman diagrams needed for $\chi^0_i\chi^0_j \to
\gamma Z$. Full internal lines denote fermionic exchanges, broken
lines with no arrows describe scalars,
and wavy lines gauge bosons. The external momenta
and the polarization vectors of the outgoing gauge bosons are
indicated in parentheses. The direction of the fermionic line is
always from $\tchi_i$ to $\tchi_j$. We call the first 10 boxes
"direct". The corresponding boxes with $\gamma\leftrightarrow Z$
exchanged, are  determined by simply enforcing (\ref{ij-antisym}),
to the above  "direct" contribution; see text. The remaining 8 boxes,
called  "twisted", automatically satisfy (\ref{ij-antisym}).}
\label{box-diag}
\end{figure}

\begin{figure}[t]
\vspace*{-1cm}
\[
\hspace{-3.cm}\epsfig{file=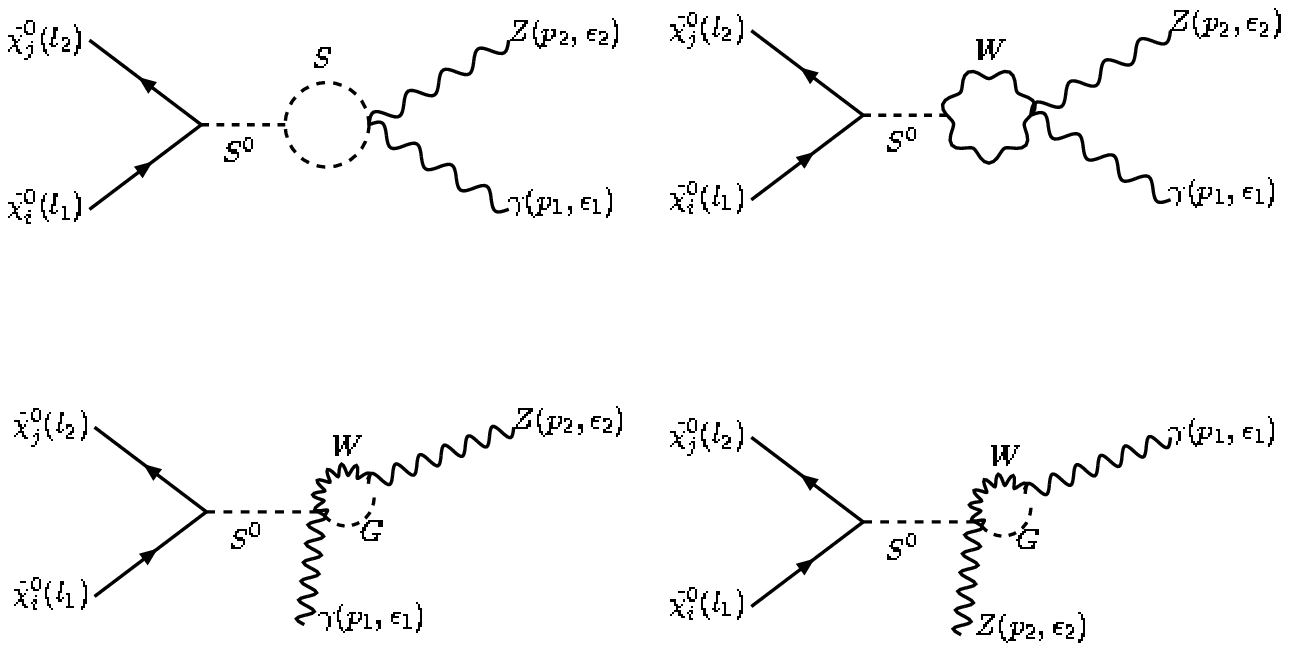,height=12.cm, width=15.5cm}
\]
\vspace{-7cm}
\[
\hspace{-3.cm}\epsfig{file=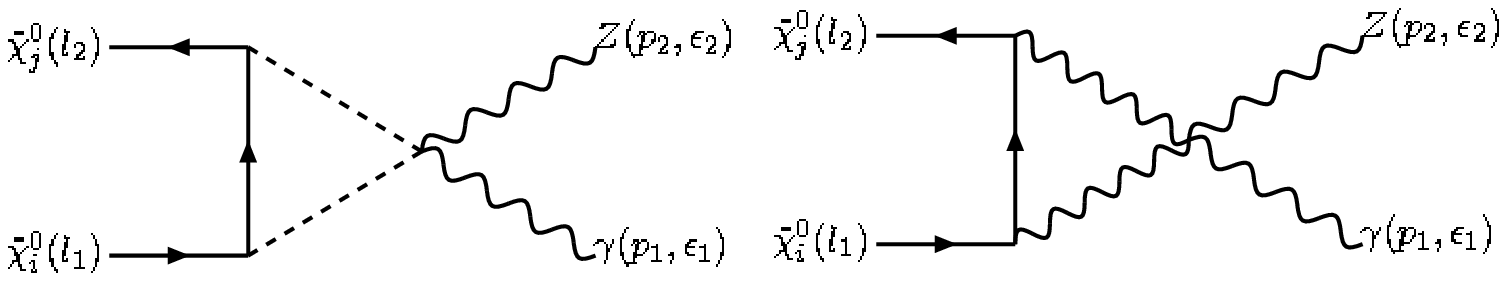,height=15.cm, width=15.5cm}
\]
\vspace{-13cm}
\[
\hspace{-0.cm}\epsfig{file=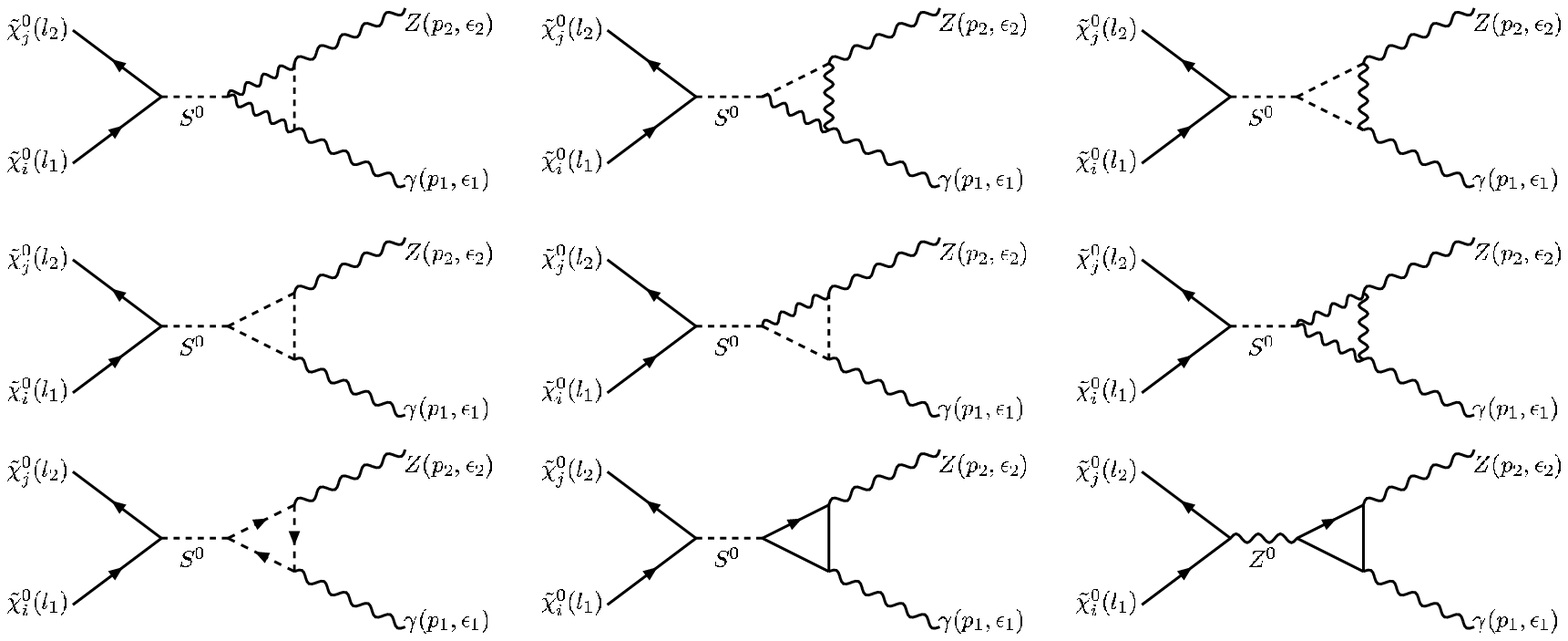,height=17.cm, width=15.5cm}
\]
\vspace{-10cm}
\caption[1]{The bubble, initial and final triangle diagrams for
$\chi^0_i\chi^0_j \to \gamma Z$. The exchanges are as in
Fig.\ref{box-diag}, but some specific  exchanges of $W$ and
goldstone bosons $G$ in the bubbles, are indicated explicitly.
$S^0$ denotes  $h^0$ or  $H^0$, except in the middle
diagram of the last line where it also describes the exchanges of
 $A^0$ and $G^0$.  $S$ denotes a charged scalar particle exchange.
The left  triangular graph  in the last line
describes the  FP ghost contribution. These contributions
satisfy (\ref{ij-antisym}).} \label{other-diag}
\end{figure}

\begin{figure}[t]
\vspace*{-1cm}
\[
\hspace{-0.cm}\epsfig{file=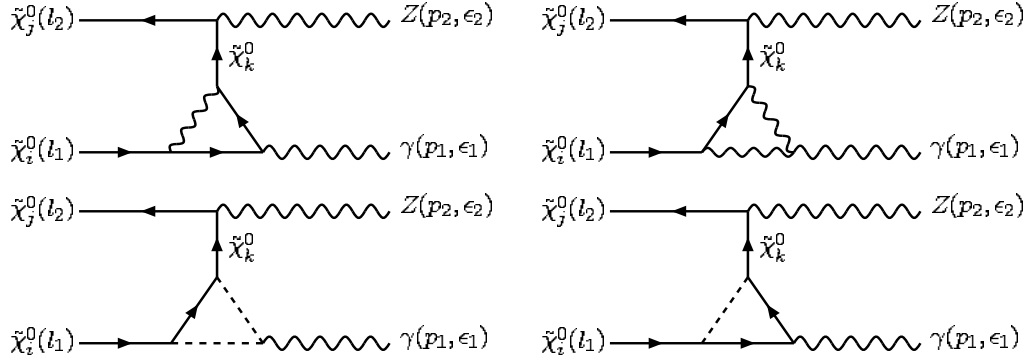,height=15.cm, width=15cm}
\]
\vspace*{-11cm}
\caption[1]{t-channel triangles. The exchanges are as in
Fig.\ref{box-diag}. The u-channel triangle contribution
is obtained from it through (\ref{u-channel-tri}); see text. } \label{t-tri-diag}
\end{figure}

\begin{figure}[b]
\vspace*{-2cm}
\[
\hspace{-0.5cm}\epsfig{file=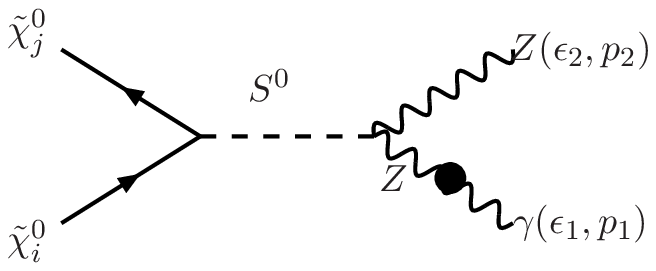,height=3.cm}
\hspace{1.cm}\epsfig{file=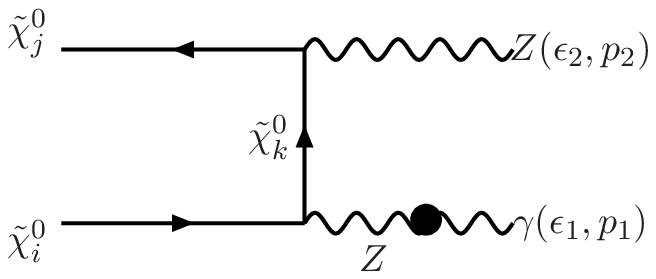,height=3.cm}
\]
\vspace*{-0.5cm}
\caption[1]{The needed $\gamma-Z$ self energy diagrams. The left
diagram   describes the s-channel exchange of $S^0=
h^0,~H^0$. The right diagram describes the   t-channel neutralino
exchange contribution, while the
  corresponding  u-channel  contribution is obtained from it
   through (\ref{u-channel-self}).}
\label{self-diag}
\end{figure}

The complete set of needed  diagrams
consists of: the box diagrams of Fig.\ref{box-diag}, the bubbles and the
initial and final triangles of Fig.\ref{other-diag},  the
t-channel triangle diagrams of Fig.\ref{t-tri-diag}, and the
$\gamma-Z$ self energy contributions\footnote{We
thank F. Boudjema for drawing our attention to this contribution}
 in  Fig.\ref{self-diag}. In all cases,
the  full internal lines denote fermionic exchanges, the broken
lines scalars, and  the wavy ones gauge bosons, while  the
arrowed broken lines   denote the usual FP
ghosts\footnote{ The only  diagram of this type appears in the
last line of in Fig.\ref{other-diag}.}. The external
momenta and the polarization vectors of the outgoing gauge bosons
are indicated in parentheses, in the figures.

Taking advantage of the Majorana nature of the neutralinos,
 the direction of the fermionic line  is always chosen
 from $\tchi_i$ to $\tchi_j$; so that the wave functions
of $\tchi_i$ ($\tchi_j$) are respectively
described by  positive (negative) energy Dirac solutions.

We next enumerate the main steps of the calculation.

We call the first 10 boxes in Fig.\ref{box-diag}    "direct". The
contribution of the corresponding boxes
with $\gamma\leftrightarrow Z$ exchanged,  is then
determined from them, by enforcing (\ref{ij-antisym}). Therefore, to
\bq
F^{ij}_{\lambda_1,\lambda_2; \mu_1,\mu_2}(\theta)^{\rm direct~ Box}~ ~,
\label{direct-box}
\eq
we should add
\bq
(-1)^{(\mu_1-\mu_2)}F^{ji}_{\lambda_2,\lambda_1;
\mu_1,\mu_2}(\pi -\theta)^{\rm direct ~ Box}~ , \label{crossed-box}
 \eq
 to take into account   the $(\gamma-Z)$-crossing contribution;
 thereby   greatly facilitating  the computation.

The last 8 boxes in Fig.\ref{box-diag}, called   "twisted", have
been checked to  satisfy (\ref{ij-antisym}) by themselves. This is
also true for the
 bubble and initial and final triangle diagrams of Fig.\ref{other-diag}.

The third  set of needed  diagrams    consists of the t-channel
triangles of Fig.\ref{t-tri-diag}, whose contribution we denote as
\bq
F^{ij}_{\lambda_1,\lambda_2; \mu_1,\mu_2}(\theta)^{\rm
t-triangle}~. \label{t-channel-tri}
\eq
To this we should add  the contribution of
the corresponding  u-channel  triangles,  obtained by
interchanging $\tchi_i \leftrightarrow  \tchi_j$, (with the arrows
always running from $\tchi_i$ to $\tchi_j$) in Fig.\ref{t-tri-diag}.
By enforcing (\ref{ij-antisym}), these u-channel triangle contribution
 is given by
\bq
F^{ij}_{\lambda_1,\lambda_2; \mu_1,\mu_2}(\theta)^{\rm u-triangle}=
(-1)^{(\mu_1-\mu_2)}F^{ji}_{\lambda_2,\lambda_1;
\mu_1,\mu_2}(\pi -\theta)^{\rm t-triangle}. \label{u-channel-tri}
 \eq

Finally, the needed $\gamma-Z$ self energy contribution is is shown in
Fig.\ref{self-diag}, where the left diagram  describes the contribution
due to s-channel exchange of\footnote{This contribution automatically satisfies
(\ref{ij-antisym}).} $(H^0,h^0)$, while the right one  describes the
contribution induced by a t-channel neutralino exchange. Associated to this
t-channel self-energy contribution, there exist a corresponding u-channel one
obtained by enforcing  (\ref{ij-antisym}) on it, through
\bq
F^{ij}_{\lambda_1,\lambda_2; \mu_1,\mu_2}(\theta)^{\rm u-{\rm self~ energy}}=
(-1)^{(\mu_1-\mu_2)}F^{ji}_{\lambda_2,\lambda_1;
\mu_1,\mu_2}(\pi -\theta)^{\rm t-{\rm self~ energy}}. \label{u-channel-self}
 \eq

 To summarize, the complete helicity amplitudes are given by the sum of the
 contributions of the diagrams in Figs.\ref{box-diag}-\ref{self-diag},
  together with the contributions appearing  in
(\ref{crossed-box}, \ref{u-channel-tri}, \ref{u-channel-self} ).  \\

As already said,
the $F^{ij}_{\lambda_1,\lambda_2; \mu_1,\mu_2}(\theta)$
amplitudes are expressed analytically in terms of the PV functions.
We  have made several tests on these results, in
order to eliminate, as much as possible, the possibility of
errors. These, we enumerate below:

Requiring  (\ref{ij-antisym}) for the contributions of
each of the 8 twisted boxes in Fig.\ref{box-diag} and of the diagrams  of
Fig.\ref{other-diag}, already provides  generally stringent tests
of their correctness.

Moreover, for the  10 direct boxes and  the first four twisted
boxes in Fig.\ref{box-diag},
which are similar to the box diagrams contributing to the
 $\gamma \gamma$-amplitudes \cite{DMgg},  we have   checked that the
 $\gamma Z$ diagrams  smoothly go  to the $\gamma\gamma$ ones,
 as $p_2^2\to 0$ and  the $Z$ couplings are replaced
 by the photon ones.

 The calculation of   the 10 direct boxes of
Fig.\ref{box-diag}, and  of the  t-channel triangles of
Fig.\ref{t-tri-diag}, for which no symmetry constraint
is available\footnote{For the $\gamma \gamma$-case, checking the 10 direct
boxes  was helped by requiring the photon-photon Bose symmetry \cite{DMgg},
which is not available here.},  has been checked several  times.

Finally, we have also  checked that our results respect
 the correct helicity conservation (HC) properties at
 high energy and  fixed angles  \cite{asymHC}.
Such tests  check stringently
 the  mass-independent high energy contributions for both
 transverse-$Z$ and longitudinal-$Z$ amplitudes.
 Any seemingly innocuous misprint,  could not only  violate HC,
 but also transform the  expected logarithmic energy dependence  of the dominant
  amplitudes at high energy, to  a linear or quadratic
  rising with energy,  thereby supplying a clear signal  of  error.
We have had ample experience of this, during our checks.

In addition to these, we have, of course,   assured  that
all  UV divergences\footnote{It is amusing to remark that
if $\Delta \equiv 1/\epsilon -\gamma+\ln(4\pi)$ is replaced by zero,
as is done by default in \cite{looptools}, and the dimensional regularization
scale is chosen as $\mu_{\rm dim}=\mw$, then the $\gamma-Z$ self energy
contribution vanishes at the 1-loop level. In this case,
the diagrams in Fig.\ref{self-diag} may be ignored.},
as well as any scale dependence, cancel out
exactly in the  amplitudes, and that  (\ref{CP-sym}) is respected,
for real MSSM parameters.\\

We next turn to the   quantities needed for DM studies of the process
(\ref{process-gZ}). These  are expressed in terms of
the  helicity amplitudes as
\bq
\Sigma_{ij}\equiv  v_{ij}\cdot \sigma_{ij}
\simeq  \frac{s-\mzd}{64 \pi [s^2-(\mid-\mjd)^2]}
\Bigg \{\int_{-1}^{+1}
d\cos\theta \sum_{\lambda_1\lambda_2\mu_1\mu_2}
|F_{\lambda_1\lambda_2\mu_1\mu_2}^{ij}|^2 \Bigg \}_s ~~,
\label{Sigmaij}
\eq
where $v_{ij}$ describes the relative velocity of the
$\tchi_i\tchi_j$-pair, implying
\bqa
s \simeq   & \equiv & (m_i+m_j)^2+m_im_j\Bigg [ v_{ij}^2 + v_{ij}^4
\Big ( \frac{3}{4}-\frac{2\mi \mj}{(\mi+\mj)^2}\Big ) \nonumber \\
&+& v_{ij}^6 ~ \frac{5(\mi^4+\mj^4)-12 \mi\mj(\mi^2+\mj^2)
+22 \mi^2\mj^2}{8(\mi+\mj)^4} \Bigg ] ~~, \label{s-v-expansion}
\eqa
up to $\Ocal(v_{ij}^6)$ terms. A numerical search indicates that
 for $v_{ij}\lsim 0.7$, the terms up to  $\Ocal(v_{ij}^4)$ in
(\ref{s-v-expansion}) should be adequate.

The transverse part  $\Sigma_{ij}^{TT}$ of $\Sigma_{ij}$
is obtained by discarding the $\mu_2=0$ (longitudinal Z)  contributions in
 the helicity summation in (\ref{Sigmaij}). \\

\section{Results and comparisons}

For understanding DM observations from neutralino-neutralino
annihilation to $\gamma Z$ in \eg the center of our Galaxy or in
nearby galaxies like Draco \cite{DMobs}, the quantity
$\Sigma_{ij}$ (\ref{Sigmaij}) should  be known  at $v_{ij}\simeq
10^{-3}$ \cite{Kamio-rep}. At so  small   velocities,  the relative orbital
angular momentum of the $\tchi_i\tchi_j$-pair must vanish, and the
system must be in either an $^1S_0$ or an $^3S_1$ state. In such
cases, the angular distribution of $d \Sigma_{ij}/d \cos\theta $
is flat.

Angular momentum
conservation implies that the $^1S_0$ state can
only contribute to transverse helicities  for  both $\gamma$
and $Z$; while the $^3S_1$ state can also give
non-vanishing longitudinal-Z contributions.

\begin{table}[pt]
{ Table 1: Results for $\Sigma_{ij}\equiv v_{ij}\sigma(\tchi_i\tchi_j\to \gamma Z)$
  at $v_{ij}=0$ and $v_{ij}=0.5$, summing over all $\gamma Z$ polarizations.
The  transverse  $Z$  contributions $\Sigma_{ij}^{TT}$ are also indicated
 in parentheses, for the relevant cases of either $i\neq j$ or $v_{ij}\neq 0$.
  Previous  results for  $i=j=1$
 from \cite{Bergstrom-gZ} at $v_{11}=0$, and  from   \cite{Boudjema} at
$v_{11}=0$ and $v_{11}=0.5$, are also compared. Inputs are at the
electroweak scale using the model sample of \cite{Boudjema}, with
 $\tan \beta=10$, and $A_f=0$ apart from
$A_t=-0.3TeV$ for $m_{\tilde f_{L,R}}=0.8TeV$ and $A_t=0$ for
$m_{\tilde f_{L,R}}=10TeV$; all  masses in TeV.}\\
  \vspace*{-1.cm}
\begin{center}
\hspace*{-0.3cm}
\begin{small}
\begin{tabular}{||c|c|c|c|c|c|c||}
\hline \hline
 & Sugra & nSugra & higgsino-1 & higgsino-2  & wino-1 & wino-2  \\ \hline
 $M_1$ & 0.2 & 0.1 & 0.5 & 20 & 0.5 & 20.0  \\
$M_2$ & 0.4 &0.4 & 1.0 & 40.&  0.2& 4.0  \\
$\mu$ & 1.0 & 1.0 & 0.2 & 4.0&  1.0& 40.0  \\
$M_A $ & 1.0 &1.0 & 1.0 & 10.0 & 1.0 & 10.0  \\
$m_{\tilde  f}$ & 0.8 & 0.8 & 0.8 & 10.0 & 0.8 & 10.0  \\
  \hline \hline
\multicolumn{7}{||c||}{$v_{11}\sigma (\tchi_1 \tchi_1 \to \gamma Z)$
   in units of    $10^{-27} cm^3 sec^{-1}$;  $v_{11} =0 $} \\
  \hline
  & $2.01\cdot 10^{-5}$  &$2.60 \cdot 10^{-6}$  &0.224
& 0.0266 & 12.6 & 10.1   \\
\cite{Boudjema} & $2.03 \cdot 10^{-5}$  & $2.61\cdot 10^{-6}$
 & 0.219  &0.0220  &11.7  & 10.1   \\
\cite{Bergstrom-gZ} & $1.42 \cdot 10^{-5}$ & $1.79 \cdot 10^{-6}$  & 0.261
 & 0.0329 & 11.7 & 10.1    \\
 \hline
 \multicolumn{7}{||c||}{ $v_{11} =0.5 $ } \\
  \hline
  & $2.45 \cdot 10^{-5}$  &$5.28 \cdot 10^{-6}$  &0.307
& 0.0165 & 14.2 & 0.575   \\
      & $(2.03 \cdot 10^{-5})$  & $(5.16 \cdot 10^{-6})$  &(0.307)
& (0.0165) & (14.2) & (0.575)   \\
\cite{Boudjema} & $2.45 \cdot 10^{-5}$  & $3.67\cdot 10^{-6}$
 & 0.299  &0.0166  &14.2  & 0.576   \\
 \hline \hline
\multicolumn{7}{||c||}{ $v_{ij}\sigma (\tchi_i \tchi_j \to \gamma Z )$
   in units of    $10^{-27} cm^3 sec^{-1}$;  $v_{ij}=0$} \\
  \hline
$\tchi_1 \tchi_2$  & $2.63 \cdot 10^{-4}$
& $9.64 \cdot 10^{-5}$  & $1.11 \cdot 10^{-3}$   & $3.28 \cdot 10^{-5}$   &
  $2.57 \cdot 10^{-3}$ & $1.04 \cdot 10^{-5}$    \\
   & $(1.94  \cdot 10^{-4})$  & $(6.02 \cdot 10^{-5})$   & $(4.12 \cdot 10^{-5})$  &
  $(1.81 \cdot 10^{-8})$  & $(1.07 \cdot 10^{-3})$ & $(8.65 \cdot 10^{-7})$     \\
$\tchi_1 \tchi_3$  & $4.51 \cdot 10^{-3}$  & $4.46 \cdot 10^{-3}$  &
 $1.59 \cdot 10^{-2}$  & $4.53 \cdot 10^{-4}$  & 0.632  & 2.90    \\
 & $(1.06 \cdot 10^{-4})$  &  $(1.06 \cdot 10^{-4})$  &
  $(9.93 \cdot 10^{-3})$   &$(2.59 \cdot 10^{-7})$  & $(6.25 \cdot 10^{-3})$
   & $(1.24 \cdot 10^{-6})$    \\
\hline
\multicolumn{7}{||c||}{ $v_{ij}=0.5 $} \\
 \hline
$\tchi_1 \tchi_2$  & $2.78 \cdot 10^{-4}$ & $1.02 \cdot 10^{-4}$
&$9.11 \cdot 10^{-4}$
 & $1.45 \cdot 10^{-8}$    & $2.45 \cdot 10^{-3}$  & $9.01 \cdot 10^{-7}$     \\
 & $(2.06 \cdot 10^{-4})$  & $(6.15 \cdot 10^{-5})$   & $(3.53 \cdot 10^{-5})$  &
  $(3.98 \cdot 10^{-11})$ & $(1.05 \cdot 10^{-3})$ & $(8.93 \cdot 10^{-7})$    \\
$\tchi_1 \tchi_3$  &$4.32 \cdot 10^{-3}$   &$4.29 \cdot 10^{-3}$   &
  $1.54 \cdot 10^{-2}$ & $3.84 \cdot 10^{-4}$ & 1.44  &  $5.56 \cdot 10^{-2}$  \\
    &$(1.10 \cdot 10^{-4})$   & $(8.34 \cdot 10^{-5})$  &
   $(9.72 \cdot 10^{-3})$ & $(2.25 \cdot 10^{-7})$  & $(4.39 \cdot 10^{-3})$
   &   $(1.31 \cdot 10^{-7})$   \\
\hline \hline
\end{tabular}
 \end{small}
\end{center}
\end{table}

If the two neutralinos happen to be  identical,
Fermi statistics only allows the $^1S_0$-state  at vanishing relative
velocities, so that  $\gamma$ and $Z$, are both  transverse.
At higher velocities though,
 like \eg   $v_{ii}=0.5$, longitudinal-Z amplitudes
can also  arise.

 On the other hand, if  $i\neq j$, longitudinal-Z amplitudes
can  also contribute,
even for vanishing relative velocities.

We also note that  for $i=j$,
the angular structure  of $d \Sigma_{ij}/d\cos\theta$,
at non-negligible  relative velocities, is  always
forward-backward symmetric. But for $i\neq j$, this is
not true any more. Depending on the content of the two
neutralinos, $d \Sigma_{ij}/d\cos\theta$ is sometimes peaked in
the forward region, and others in the backward;
compare (\ref{ij-antisym}). For $v_{ij}\lsim 0.5$,
$d \Sigma_{ij}/d\cos\theta$ was found to be flat,
in all examples we have considered.\\

Next, we turn to the specific features of our approach based on the PV functions,
whose definition is known to be singular close to threshold and at the forward or
backward angles\footnote{These singularities are solely due to the
mathematical definitions used. They do not have anything to do
with the physical problem and they do not appear in the total
amplitude.} \cite{looptools}.
Thus, for  relative velocities  of $v_{ij}\simeq 10^{-3}$,
or for angles in the forward or backward region, extrapolations
must be done. In all examples we have considered,
these were very smooth, with no suggestion of a possible introduction of errors.

The  herewith released code PLATONdmZ  calculates
  $d \Sigma_{ij}/d\cos\theta$ in fb \cite{plato},
  for real MSSM parameters at the
  electroweak scale, and   fixed values of the relative
  neutralino velocity $v_{ij}$, and\footnote{To transform
it  to the usual DM units of $cm^3/sec$ we should multiply it by $
3 \times 10^{-29}$.} $\cos\theta$ using  \cite{looptools}.
For  $0.1\lsim v_{ij} \lsim 0.7$
 and angles away from the forward and backward regions, the
 direct use of the code  usually runs without problems.

The only known exception  appears in cases where the sum of the
masses of the two annihilating neutralinos happens to be close to
the Z-pole. Such additional threshold singularities are specific
for the $\tchi_i\tchi_j\to \gamma Z$  mode, and they
have no counterpart in  the
corresponding $\gamma \gamma$- and $gg$ calculations of  \cite{DMgg}.\\

The next step is to compare  our work,
 with that  of other authors'. The
only preexisting results in the literature apply to i=j=1 for
$v_{11}=0$ and $v_{11}=0.5$, presented by    \cite{Boudjema}, and
for $v_{11}=0$ presented by  \cite{Bergstrom-gZ}. To compare with
them,  we  give  in Table 1 the results of the PLATONdmZ code,
together with those of \cite{Boudjema} and \cite{Bergstrom-gZ}.
We use  the same sample of models as in\footnote{We also use
 $\mz=0.091187TeV$ and $\swd=0.2319$, as in \cite{Boudjema},
 and $m_t=0.174 TeV$. } \cite{Boudjema}.
  These models, whose electroweak scale parameters are
  indicated in  Table 1, have been named  Sugra, nSugra,
higgsino-1, higgsino-2, wino-1 and wino-2 by \cite{Boudjema}.
In Sugra and nSugra, the lightest neutralino (LSP) is a bino.
In higgsino-1 and higgsino-2, the two lightest neutralinos are
 almost or exactly degenerate higgsinos. Finally,
in  wino-1 and wino-2, the lightest supersymmetric particle (LSP)
is a wino, while  the NLSP is a bino.

As seen in Table 1,  the  PLATONdmZ results for i=j=1 and  $v_{11}=0$,
perfectly agree with those of \cite{Boudjema}, while they deviate
from those of \cite{Bergstrom-gZ}.
As expected,     $Z$   is completely  transverse, in this case.

For  $v_{11}=0.5$ though, longitudinal-Z contributions, are also
possible. Because of this, in Table 1, we first give the results
for the full  $\gamma Z$-production, while in  parentheses, in
the next line, the completely transverse $\gamma Z$ production is also
indicated. As shown in this Table, appreciable longitudinal Z
contribution at $v_{11}=0.5$, only appears for Sugra.

 At $v_{11}=0.5$,
important discrepancies between our predictions for  $\Sigma_{11}$ and those
of \cite{Boudjema}, only appear  for nSUGRA, reaching    the 40\%
 level.

We also note that in nSugra and wino-2, $\Sigma_{11}$ is very
sensitive to  the relative velocity $v_{11}$. This can be inferred
from the big difference  between the $v_{11}=0$ and
$v_{11}=0.5$ results, for $i=j=1$, in  these models.
For nSugra this is also
elucidated in Table 2. Is this sensitivity partly responsible for
the  discrepancy in Table 1, of the present results,
with respect to those of  \cite{Boudjema}? And then,
why it does  not induce any discrepancy in wino-2?\\

\begin{table}[ht]
\begin{center}
{ Table 2: Sensitivity of $\Sigma_{11}\equiv  v_{11}\cdot \sigma_{11}$
to $v_{11}$ in nSugra. }\\
  \vspace*{0.3cm}
\begin{small}
\begin{tabular}{||c|c||}
\hline \hline
  \multicolumn{2}{||c||}{nSugra }\\
  \hline
$v_{11}$ &   $\Sigma_{11}$ $(10^{-27} cm^3 sec^{-1})$  \\
\hline
0.09 & $2.65 \cdot 10^{-6}$  \\
 0.2 &  $3.03 \cdot 10^{-6}$  \\
 0.3 &   $3.61 \cdot 10^{-6}$ \\
 0.5 &   $5.28 \cdot 10^{-6}$ \\
 \hline \hline
\end{tabular}
 \end{small}
\end{center}
\end{table}

In Table 1 we also give results for $\Sigma_{12}$ and
$\Sigma_{13}$  for the above  6 models, at $v_{ij}=0$ and
$v_{ij}=0.5$. In parentheses, the purely transverse $\gamma Z$
contributions are also indicated, which are generally not identical
to the total cross sections. Longitudinal Z production
 is often  important in these cases, and sometimes it even dominates
the transverse Z contribution, at both shown velocities.

  It is also amusing to notice from Table 1
  the sensitivities of $\Sigma_{12}$ and $\Sigma_{13}$ on $v_{ij}$,
    as it changes from 0.0 to 0.5, in  the various models.
    For \eg  $\Sigma_{12}$, strong  sensitivity  appears
    in the higgsino-2 and wino-2 cases; while for  $\Sigma_{11}$
    a corresponding phenomenon is observed for nSUGRA and  wino-2 again. \\

\section{Conclusions and Outlook}

 The neutralinos may be the most abundant particles in the
Universe, if they  really turn out to contribute appreciably to its Dark
Matter. They are thus,  very interesting objects. In addition to
this,  they are very interesting from the particle physics point of view,
since their Majorana nature allows them to interact
in many  more ways, than the ordinary (neutral) Dirac fermions.
Because of this,  detailed studies  of their properties,
both, in astrophysical observations and accelerator experiments  are welcomed.

Through the present paper, an extensive analytical study of
the 1-loop neutralino amplitudes in any unconstrained
 minimal supersymmetric model (MSSM) with real parameters,
 has been completed, and the related FORTRAN
codes have been released  \cite{plato}.

More explicitly, the DM relevant process
\bq
 \tchi_i \tchi_j \to \gamma  Z ~ ,  \label{DMnngZ}
\eq
has been studied analytically here, for any kinematic configuration, while
\bq
 \tchi_i \tchi_j \to \gamma \gamma~, \label{DMnngg}
\eq
has been presented in \cite{DMgg}, following  the same spirit.
The reverse process
$\gamma \gamma  \to \tchi_i \tchi_j $, which is suitable for a $LC_{\gamma\gamma}$
collider study, has appeared in \cite{LCggnn};
while the LHC production processes, containing  two or one neutralino,
 have appeared  in \cite{LHCnn} and \cite{LHCn} respectively.

The formalism of all these 1-loop processes is
quite common\footnote{To the LHC studies \cite{LHCnn, LHCn},
processes receiving tree level contributions also appear.
But the expressions for them are so simple, that no codes are needed.},
while the couplings are, of course,  different in each case.
Thus, the 1-loop Feynman diagrams  determining  the  amplitudes
for neutralino production at LHC, constitute a subset of those entering
$\tchi_i \tchi_j \to \gamma \gamma$ production, which in turn
comprise a subset of those needed for  $\tchi_i \tchi_j \to \gamma Z$.

The latter,  determines also the "reverse" process \bq
 \gamma + e^\mp  \to \tchi_i +\tchi_j +e^\mp~,
 \label{LCegnn}
\eq
where an off-shell intermediate  $\gamma$ or $Z$,
emitted by the incoming $e^\mp$-line, interacts
with another incoming photon, producing  a pair of neutralinos.
Such a process could be studied in a future
$e^\mp\gamma$ Linear Collider, providing  further constraints on
neutralinos and DM.  It is straightforward
to get the amplitudes for (\ref{LCegnn}), from those of the  process
(\ref{DMnngZ}), studied here.
We hope to present results for this  in the  future.

It would be  really thrilling,  if we  ever unambiguously identify
energetic $\gamma$-rays coming from  Space
and being associated to  Dark Matter annihilation
\cite{DMobs}. It would be even more so, if, along with the
continuous $\gamma$-spectrum, we could also detect  the
sharp monochromatic photons implied by (\ref{DMnngg},
\ref{DMnngZ}). But even if this turns out to be the
case, the neutralino DM interpretation  will not be sufficiently convincing,
unless detailed accelerator studies confirm it.
The   present  work     contributes towards  this.

\vspace{0.4cm}
\noindent
\underline{Acknowledgments}:\\
We are grateful to Fawzi Boudjema for important
remarks related to the comparison
of our work with that of ref.\cite{Boudjema}.
 GJG  gratefully acknowledges also the  support from the
 European Union program    MRTN-CT-2004-503369.

\clearpage
\newpage

\clearpage
\newpage

\clearpage
\newpage

\newpage

\begin{thebibliography}{99}

%
\bibitem{Kamio-rep}D.N. Spergel \etal arXiv:astro-ph/0302209;
 G. Jungman, M. Kamionkowski and K. Griest,
\prep{267}{195}{1996}; M. Kamionkowski, hep-ph/0210370;
M. Drees, Pramana {\bf 51},87(1998);
M.S. Turner, J.A. Tyson, astro-ph/9901113, \rmp{71S}{145}{1999};
M.M. Nojiri, hep-ph/0305192; M. Drees hep-ph/0210142;
 D.P. Roy, \polon{B34}{3417}{2003},
hep-ph/0303106; F.E. Paige, hep-ph/0307342, hep-ph/0211017;
J.A. Aguilar-Saavedra \etal, SPA project, hep-ph/0511344;
D. Fargion, R. Konoplich, M. Grossi and M.Yu.Khlopov,
\app{12}{307}{2000}, astro-ph/9902327.
%
\bibitem{Laz} For a recent review see \eg G. Lazarides, hep-ph/0601016.
%
\bibitem{SUSYsearches}G. Bertone, D. Hooper, J. Silk,
\prep{405}{279}{2005}, hep-ph/0404175; J.L. Feng, hep-ph/0405215.
%
\bibitem{Bergstrom-gg}
L. Bergstr\"om and P. Ullio, \np{B504}{27}{1997}, hep-ph/9706232;
 Z. Bern, P. Gondolo and M. Perelstein, \pl{B411}{86}{1997},
hep-ph/9706538].
%
\bibitem{Bergstrom-gZ}
 P. Ullio and L. Bergstr\"om, \pr{D57}{1962}{1998}.
%
\bibitem{DMgg} G.J. Gounaris, J. Layssac,
P.I. Porfyriadis and F.M. Renard, \pr{D69}{075007}{2004},
hep-ph/0309032.
%
\bibitem{Veltman} G. Passarino and M. Veltman \np{B160}{151}{1979}.
%
\bibitem{LCggnn} G.J. Gounaris, J. Layssac,
P.I. Porfyriadis and F.M. Renard, \epj{C32}{561}{2004},
hep-ph/0311076.
%
\bibitem{LHCnn} G.J. Gounaris, J. Layssac,
P.I. Porfyriadis and F.M. Renard, \pr{D70}{033011}{2004},
hep-ph/0404162.
%
\bibitem{LHCn} G.J. Gounaris, J. Layssac,
P.I. Porfyriadis and F.M. Renard, \pr{D71}{075012}{2005},
hep-ph/0411366.
%
\bibitem{plato} PLATON codes can be downloaded
from http://dtp.physics.auth.gr/platon/
%
\bibitem{Boudjema}F. Boudjema, A. Semenov and D. Temes,
\pr{D72}{055024}{2005}, hep-ph/0507127.
%
\bibitem{JW}M. Jacob and G.C. Wick, \aop{7}{404}{1959},
\aop{281}{774}{2000}.
%
\bibitem{LeMouel1}See \eg G.J. Gounaris, C. Le Mou\"{e}l and
P.I. Porfyriadis, \pr{D65}{035002}{2002}, hep-ph/0107249.
%
\bibitem{asymHC} G.J. Gounaris and F.M. Renard, \prl{94}{131601}{2005},
hep-ph/0501046; G.J. Gounaris hep-ph/0510061;
G.J. Gounaris and F.M. Renard, in preparation.
%
\bibitem{DMobs} W. de Boer, hep-ph/0508108;
L. Bergstr\"om and D. Hooper, hep-ph/0512317;
S. Profumo and M. Kamionkowski, astro-ph/0601249;
Y. Mambrini and E. Nezri, hep-ph/0507263;
Y. Mambrini, C. Mu\~{n}oz, E. Nezri and F. Prada, hep-ph/0506204.
%
\bibitem{looptools} T. Hahn, LoopTools,
http://www.fsf.org/copyleft/lgpl.html;
T. Hahn and M. P\'{e}rez-Victoria, hep-ph/9807565;
G.J. van Oldenborgh and J.A.M. Vermaseren,
\zp{C46}{425}{1990}.
%




\end{thebibliography}
\end{document}